\documentclass[prd,aps,notitlepage,superscriptaddress,nofootinbib]{revtex4-1}
\usepackage{amsmath}
\usepackage{slashed}
\usepackage{epsfig}
\usepackage{color}
\def\bea#1\eea{\begin{align}#1\end{align}} 
\newcommand{\nnu}{\nonumber\\}
\newcommand{\bef}{\begin{figure}[htb]\centering}
\newcommand{\eef}{\end{figure}}

\def\taun{{\cal T}_N}

\def\tauncut{{\cal T}_N^{cut}}

\allowdisplaybreaks

\begin{document}
\title{The spin-dependent quark beam function at NNLO}

\date{\today}

\author{Radja Boughezal}
\email{rboughezal@anl.gov}
\affiliation{High Energy Physics Division, Argonne National Laboratory, Argonne, Illinois 60439, USA}

\author{Frank Petriello}
\email{f-petriello@northwestern.edu}
\affiliation{High Energy Physics Division, Argonne National Laboratory, Argonne, Illinois 60439, USA}
\affiliation{Department of Physics and Astronomy, Northwestern University,  Evanston, Illinois 60208, USA}

\author{Ulrich Schubert}
\email{schubertmielnik@anl.gov}
\affiliation{High Energy Physics Division, Argonne National Laboratory, Argonne, Illinois 60439, USA}

\author{Hongxi Xing}
\email{hxing@northwestern.edu}
\affiliation{High Energy Physics Division, Argonne National Laboratory, Argonne, Illinois 60439, USA}
\affiliation{Department of Physics and Astronomy, Northwestern University,  Evanston, Illinois 60208, USA}

\begin{abstract}
We calculate the beam function for longitudinally-polarized quarks through next-to-next-to-leading order (NNLO) in QCD perturbation theory.  This is the last missing ingredient needed to apply the factorization theorem for the $N$-jettiness event-shape variable in polarized collisions through the NNLO level.  We present all technical details of our derivation.    As a by-product of our calculation we provide the first independent check of the previously-obtained unpolarized quark beam function.  We anticipate that our result will have phenomenological applications in describing data from polarized collisions.

\end{abstract}

\maketitle

\section{Introduction}

The precision of modern hadron collider experiments has reached an impressive level.  At the Large Hadron Collider (LHC), sub-percent experimental errors have been obtained in numerous measurements that span orders of magnitude in energy.  Theoretical advances have matched the pace of this experimental progress.  New calculations and simulation tools have been developed over the past decade that have made possible precision comparisons between theory and data that reach the percent level for some measurements.  These achievements together test the high-energy predictions of the Standard Model more stringently than ever before.  

There is a desire to extend such rigorous quantitative comparisons to polarized collisions as well.  New data is becoming available for polarized collisions with direct implications for measurements of fundamental importance, such as the determination of the polarized gluon content of the proton at the Relativistic Heavy Ion Collider (RHIC)~\cite{Adamczyk:2014ozi}.  Several other experiments have released measurements of the double-spin asymmetry in polarized collisions~\cite{Anthony:1999ac,Airapetian:2010ac}.  The theoretical description of this data using perturbative QCD predictions at next-to-leading order (NLO) in the strong coupling constant is not entirely satisfactory~\cite{Hinderer:2017ntk}.  More precise data is expected at a future electron-ion collider (EIC).  The EIC is expected to be a unique laboratory in which to discover novel QCD phenomena and to determine the spin structure of the proton with an accuracy currently not achievable.  In order to achieve these goals perturbative QCD effects will need to be disentangled from new non-perturbative phenomena and proton structure.  The situation will be very similar to the current one at the LHC, where higher-order perturbative QCD effects must be distinguished from potential signals of beyond-the-Standard Model phenomena.

In order to quantitatively describe and analyze data from polarized collisions, more theoretical advances are necessary.  As mentioned, NLO perturbative QCD fails to properly describe double-spin asymmetries appearing in deeply-inelastic scattering (DIS) in kinematic regions where non-perturbative effects are expected to be suppressed~\cite{Hinderer:2017ntk}.  The situation may be compared to the case of inclusive jet production in unpolarized DIS that has received attention recently.  The NLO corrections were found to be large, increasing the leading-order result by nearly a factor of two~\cite{Hinderer:2015hra}.  The perturbative series stabilized only after the inclusion of next-to-next-to-leading order (NNLO) corrections~\cite{Abelof:2016pby}.  This suggests that calculations at this order in perturbation theory may be necessary for a proper description of current polarized hadron-collider data.  As the experimental errors are expected to decrease at an EIC, it is expected that NNLO theory will become an integral part of the EIC physics program.

One very successful approach to higher-order calculations for unpolarized collisions at the LHC has been $N$-jettiness subtraction~\cite{Boughezal:2015dva,Gaunt:2015pea}.  This method has been used to compute the NNLO corrections for a host of jet production processes~\cite{Boughezal:2015dva,Boughezal:2015dva,Boughezal:2015aha,Boughezal:2015ded,Boughezal:2016yfp,Boughezal:2016dtm,Boughezal:2016isb,Campbell:2016lzl}, as well as numerous color-singlet production processes~\cite{Boughezal:2016wmq}.  It has been extended to handle inclusive jet production in electron-nucleon collisions as well~\cite{Abelof:2016pby}.  $N$-jettiness subtraction is based on the event-shape variable $\taun$~\cite{Stewart:2010tn,Stewart:2009yx,Stewart:2010qs}:
\begin{equation}
{\cal T}_N = \sum_k \text{min}_i \left\{ \frac{2 p_i \cdot q_k}{Q_i}\right\}.
\end{equation}
The subscript $N$ denotes the number of jets desired in the final state, and is an input to the measurement. The $q_k$ denote the four-momenta of any final-state radiation, while the $p_i$ denote the momenta of the initial-state hard partons and any final-state jets.  The $Q_i$ are dimensionful variables that characterize the hardness of the beam-jets and final-state jets.  The cross section factorizes in the limit where $\taun$ is less than any other hard scale in the problem~\cite{Stewart:2010tn,Stewart:2009yx,Stewart:2010qs}.  Schematically, the form of the cross section in this limit becomes
\begin{equation} \label{eq:factschem}
 \frac{ d\sigma}{d \taun} = H \otimes B \otimes S \otimes   \left[ \prod_{n}^{N} J_n \right] +\cdots .
\end{equation}
Here, $H$ describes the effect of hard radiation, $S$ describes the soft radiation, and $J_n$ contains the radiation collinear to a final-state jet.  We note that the $J_n$ can be straightforwardly replaced by a fragmentation function if a final-state hadron is instead measured. $B$ encodes the effect of radiation collinear to an initial beam direction.  We have assumed a single hadronic beam as for DIS; this formula would contain two beam functions for proton-proton collisions, or none for $e^+e^-$ collisions.  Depending on the observable and process under consideration, only a subset of the other terms may be present.  The ellipsis denote power-suppressed terms which become negligible for $\taun \ll Q_i$.  There has been recent progress in understanding these terms for color-singlet production~\cite{Moult:2016fqy,Boughezal:2016zws,Moult:2017rpl}.  The derivation of this factorization theorem relies heavily upon the machinery of Soft-Collinear Effective Theory (SCET)~\cite{scet}.  As a NNLO calculational technique, $N$-jettiness subtraction works by partitioning the phase space using a cut $\tauncut$ with  $\tauncut \ll Q_i$, using the factorization theorem of Eq.~(\ref{eq:factschem}) below this cut, and noting that the cross section above the cut becomes a simpler NLO cross section.  We refer the reader to the original papers for more details~\cite{Boughezal:2015dva,Gaunt:2015pea}.  We note that the study of $N$-jettiness has intrinsic interest besides its use as a NNLO subtraction scheme.  It is widely used in study of jet substructure through its $N$-subjettiness incarnation~\cite{Thaler:2010tr}, and has been proposed as a measure of the nuclear medium in electron-nucleus collisions~\cite{Kang:2012zr,Kang:2013wca}.  A key aspect of the usefulness of $N$-jettiness at the LHC is our ability to calculate to high orders in the QCD perturbative expansion the objects that appear in the factorization theorem of Eq.~(\ref{eq:factschem}).  The beam function $B$~\cite{Gaunt:2014xga,Gaunt:2014cfa}, the jet function $J_n$~\cite{Becher:2006qw, Becher:2010pd} and 
the soft function ${\cal S}_N$ for jets~\cite{Boughezal:2015eha} and for the massive case~\cite{Li:2016tvb}
 are all known to the NNLO level, as are the hard functions for many processes of interest.

Given the success of this framework in describing a host of data at the LHC and at other unpolarized colliders, and the need for higher-order corrections to better describe current data from polarized collisions as well as expected future data from an EIC, it is interesting to extend the $N$-jettiness framework to cover polarized collisions.  As a concrete example we consider the double spin asymmetry in lepton-proton collisions, ${\cal A}_{LL}$.  This observable begins at the leading order in the twist expansion.  It is straightforward to write down the analogous factorization theorem in the low-$\taun$ limit for the relevant polarized cross section that enters the numerator of this asymmetry\footnote{The possible contribution of perturbative Glauber modes to this factorization theorem is expected to occur at higher orders in the strong coupling constant than the NNLO level of interest here~\cite{Zeng:2015iba}.}:
\begin{equation} \label{eq:factschempol}
 \frac{ d \sigma_{LL}}{d \taun} = \Delta H \otimes \Delta B \otimes S \otimes   \left[ \prod_{n}^{N} J_n \right] +\cdots .
\end{equation}
Here, the $\Delta$ denotes the polarization dependence caused by taking the appropriate difference of helicities needed to obtain the double-spin asymmetry.  We note that the soft function $S$ and the jet/fragmentation function $J_n$ are unchanged upon considering polarized collisions.  Furthermore, the hard function is obtained from the virtual corrections to the scattering process under consideration.  Since these are generally computed for the separate helicity states, $\Delta H$ is known for most processes of interest.  Only the polarized beam function $\Delta B$ is not known at the NNLO level.  In analogy to the beam function for unpolarized collisions, the polarized beam function is a non-perturbative object that can be matched to the polarized parton distribution functions~\cite{Stewart:2010tn,Stewart:2009yx,Stewart:2010qs}:
\begin{equation}
\label{eq:schemmatch}
\Delta B_i = \sum_j \Delta {\cal I}_{ij} \otimes \Delta f_j \left[ 1+ {\cal O}\left( \frac{\Lambda_{{\rm QCD}}^2}{\taun}\right)\right].
\end{equation}
The $ \Delta {\cal I}_{ij}$ are perturbatively calculable matching coefficients, while the $\Delta f_j $ are the standard polarized PDFs. The $i$ and $j$ are parton labels. From this expression it is also apparent that studying the polarized beam function may reveal aspects of the polarized PDFs.  

It is our goal in this manuscript to calculate the polarized quark beam function matching coefficients through the NNLO level.  This is the last missing ingredient needed to bring the theoretical status of polarized collisions at leading twist to the same level as for unpolarized collisions.  It will allow for NNLO calculations in polarized collisions, and will be a necessary ingredient in extending the global fit of polarized proton structure~\cite{deFlorian:2014yva} to the NNLO level.  The beam function can be thought of as a generalized PDF, where the operators appearing in the definition are separated along both light-cone directions.  In this sense our calculation of $\Delta B$ represents the first NNLO calculation of such an object for polarized collisions.  We discuss all relevant details of our calculation, including our treatment of $\gamma_5$ in dimensional regularization.  As a by-product of our calculation we provide the first independent check of the NNLO unpolarized quark beam function.  We find complete agreement with the previous result for this quantity~\cite{Gaunt:2014xga}.  However, we do not find any need for an additional regulator beyond dimensional regularization in our calculation, in apparent contrast to this previous result.  

Our paper is organized as follows.  We define the spin-dependent beam function and establish our notation in Section~\ref{sec:setup}.  In Section~\ref{sec:renmatch} we discuss the renormalization and matching needed to convert the bare results into the renormalized matching coefficients needed in Eq.~(\ref{eq:schemmatch}).  We present the details of our NLO calculation in Section~\ref{sec:nlo}.  The details of our NNLO derivation are presented in Section~\ref{sec:nnlo}.  Finally, we conclude in Section~\ref{sec:conc}.

\section{Setup and definition of the spin-dependent beam function}
\label{sec:setup}

We begin by establishing our notation and defining the spin-dependent beam function.  We will use the standard light-cone vectors $n^{\mu}, \bar{n}^{\mu}$ with $n^2=\bar n^2 = 0$ and $n \cdot \bar{n}=2$. Any four vector can be written in terms of these directions as $p^{\mu} = (p^+, ~ p^-, ~ p_{\perp}^{\mu}) = (n\cdot p, ~\bar n\cdot p, ~p_{\perp}^{\mu})$.

The longitudinal spin-dependent beam function represents the difference of beam functions with positive and negative helicity in a parent proton with positive helicity. Focusing on the quark and anti-quark beam functions, we can define them as the proton matrix element of quark operators:
\bea
\label{eq:opdef}
\Delta B_q(t, x, \mu) =&~ \langle p_n(P^-), + | \theta(\omega) \bar\chi_n(0)\delta(t-\omega \hat p^+)\frac{\slashed {\bar n}\gamma_5}{2}
[\delta(\omega-\overline{\mathcal{P}}_n)\chi_n(0)]|p_n(P^-), + \rangle \\
\Delta B_{\bar q}(t, x, \mu) =&~ \langle p_n(P^-), + | \theta(\omega) \frac{\slashed {\bar n}\gamma_5}{2} \chi_n(0)\delta(t-\omega \hat p^+)
[\delta(\omega-\overline{\mathcal{P}}_n)\bar \chi_n(0)]|p_n(P^-) , + \rangle.
\eea 
Here, the ``$+$" represents the positive helicity of the proton, $x=\omega/P^-$ is the momentum fraction of the proton carried by the parton that enters the hard scattering, the $\delta(\omega-\overline{\mathcal{P}}_n)$ operator constrains the total minus momentum of the composite quark/gluon field to $\omega$, and $\delta(t-\omega \hat p^+)$ sets the total plus momentum of all initial state radiations to $t/\omega$.  $\chi$ is the composite quark operator
\bea
\chi_n(y) = W_{n}^{\dagger}(y) \xi_n(y), 
\eea
where $\xi_n$ is the n-collinear quark field, and $W_n$ is the Wilson line
\bea
W_n(y) = \left[\sum_{\rm perms}{\rm exp}\left(-\frac{g}{\overline {\mathcal{P}}_n}\bar n\cdot A_n(y)\right)\right].
\eea
We refer the reader to the SCET literature for more details on the operators that appear in these definitions~\cite{scet}.  Our definition of the polarized beam function follows the definition of the polarized PDF~\cite{Manohar:1990jx} with the appearance of the additional plus momentum component $t / \omega$.  We note that $t$ represents the beam-sector contribution to the measured $N$-jettiness $\taun$.
 
As discussed in the introduction the polarized beam function is a non-perturbative quantity that can be matched to polarized PDFs, in analogy to the unpolarized beam function~\cite{Stewart:2010tn,Stewart:2009yx,Stewart:2010qs}.   In order to calculate the matching coefficients, we replace the proton state by $n$-collinear quark and gluon states with momentum $p=(0,p^-,0)$.   The desired matching coefficients are unchanged upon making this replacement.  With this substitution the matching equation takes the form 
 \bea
\label{eq:match}
\Delta B_{ij}(t,z,\mu) = \sum_k \Delta{\mathcal I}_{ik}(t,z,\mu) \otimes   \Delta f_{kj}\left(z \right)  \equiv \sum_k \int_z^1\frac{d z'}{z'} \Delta{\mathcal I}_{ik}(t,z',\mu) \Delta f_{kj}\left(\frac{z}{z'}\right).
\eea
The quantity $\Delta f_{kj}$ is the distribution function for a parton of flavor $k$ within another parton of flavor $j$ (we have replaced the proton state by $j$).  The $\Delta B_{ij}$ are the polarized beam functions with this replacement for the PDFs.   The tree level diagram for quark beam function with an external quark is
 \bea
 \Delta B_{qq}^{(0)}(t, z, \mu) = \langle q_n(p), + | \theta(\omega) \bar\chi_n(0)\delta(t-\omega \hat p^+)\frac{\slashed {\bar n}\gamma_5}{2}
[\delta(\omega-\overline{\mathcal{P}}_n)\chi_n(0)]|q_n(p), + \rangle
= \delta(t)\delta(1-\omega/p^-).
 \eea 
The tree-level matching coefficient is therefore
\bea
\mathcal{I}_{qq}^{(0)}(t,z,\mu) = \mathcal{I}_{\bar{q}\bar{q}}^{(0)}(t,z,\mu)  = \delta(t)\delta(1-z).
\eea
Inserting instead a gluon in place of the the initial-state proton leads to a vanishing result, allowing us to conclude that
\bea
\mathcal{I}_{qg}^{(0)}(t,z,\mu) = \mathcal{I}_{gq}^{(0)}(t,z,\mu) =0.
\eea

\section{Renormalization and matching}
\label{sec:renmatch}

At higher orders in the strong coupling constant we must renormalize the beam function, and also perform the matching to the PDFs.  The matching equation has already been presented in Eq.~(\ref{eq:match}).  The bare and renormalized beam functions are related through the renormalization constants $Z_i$:
\bea
\Delta B_{ij}^{bare}(t,z) = \int dt' Z_{i}(t-t',\mu) \Delta B_{ij}(t',z,\mu) \, ,
\label{B_bare}
\eea
where the bare beam function depends on the renormalized $\overline{\rm MS}$ coupling $g$, and the renormalization constants $Z_i$ are defined to remove UV divergences in the bare beam function. Taking derivatives on both sides of Eq.~(\ref{B_bare}) with respect to $\mu$, one can derive the renormalization group equations (RGEs) for the polarized beam functions:
\bea
\mu\frac{d}{d\mu}\Delta B_{ij}(t',z,\mu)=\int dt^{\prime}\gamma_B^{i}(t-t^{\prime},\mu)\Delta B_{ij}(t',z,\mu),
\eea
where the anomalous dimension for the quark beam function in the $\rm\overline{MS}$ scheme is defined as
\bea
\gamma_B^{i}(t,\mu) = -\int dt' (Z_{i})^{-1}(t-t',\mu)\mu\frac{d}{d\mu}Z_{i}(t',\mu).
\eea
The inverse of $Z_{i}$ is defined as 
\bea
\int dt' (Z_{i})^{-1}(t-t',\mu)Z_{i}(t',\mu)=\delta(t).
\eea
As we will see later from explicit calculations at NLO and NNLO, the renormalization constant is the same in the polarized and unpolarized cases. This indicates that the RGEs for polarized beam functions follow exactly the same form as in the unpolarized case.

To facilitate the expansion in the strong coupling constant we introduce separate expansions for each of the quantities that appear in our result:
\bea
\Delta B_{ij} =& \sum_n\left(\frac{\alpha_s}{4\pi}\right)^n \Delta B_{ij}^{(n)},\nnu
Z_{i} =& \sum_n\left(\frac{\alpha_s}{4\pi}\right)^n Z_{i}^{(n)},\nnu
\Delta{\mathcal I}_{ij} =& \sum_n\left(\frac{\alpha_s}{4\pi}\right)^n \Delta{\mathcal I}_{ij}^{(n)},\nnu
\Delta f_{ij} =& \sum_n\left(\frac{\alpha_s}{2\pi}\right)^n \Delta f_{ij}^{(n)}.
\eea
The different choices of two and four in these expansions match the typical conventions in the literature for the various objects.  At order $\alpha_s$, using the fact that 
\bea
 \Delta B_{ij}^{(0)}(t',z,\mu) = \delta_{ij}\delta(t')\delta(1-z),~~~Z_{i}^{(0)}(t-t',\mu) = \delta(t-t'),
\label{eq-B0Z0}
\eea
we can derive the following relation between the renormalized and bare beam functions at NLO:
\bea
\Delta B_{ij}^{bare(1)}(t,z) = \Delta B_{ij}^{(1)}(t,z,\mu) + Z_{i}^{(1)}(t,\mu)\delta_{ij}\delta(1-z).
\label{eq-z1}
\eea
Similarly, we can expand Eq.~(\ref{B_bare}) to obtain the analogous relation for the NNLO beam functions:
\bea
\Delta B_{ij}^{bare(2)}(t,z) = \Delta B_{ij}^{(2)}(t,z,\mu) + Z_{i}^{(2)}(t,\mu) \delta_{ij}\delta(1-z)
+\int dt' Z_{i}^{(1)}(t-t',\mu)  \Delta B_{ij}^{(1)}(t',z,\mu).
\label{eq-z2}
\eea
We will use standard $\overline{{\rm MS}}$ renormalization, so that the $Z_i^{(n)}$ renormalization constants will contain only poles in the dimensional regularization parameter $\epsilon=(4-d)/2$, where $d$ is the space-time dimension. 

An important technical issue to discuss is the treatment of $\gamma_5$ in $d$-dimensions.  Several consistent schemes have been proposed for this purpose.  We use the HV scheme~\cite{tHooft:1972tcz,Breitenlohner:1977hr}, in which $\gamma_5$ maintains its 4-dimensional definition: $\gamma_5 = i \gamma^0 \gamma^1 \gamma^2 \gamma^3$.  Denoting 4-dimensional quantities with a tilde and $\epsilon$-dimensional ones with a hat, this definition leads to the following commutation and anti-commutation rules for $\gamma_5$:
\begin{equation} 
\label{eq:HVBM}
\{ \gamma_5,\tilde{\gamma}_{\mu} \} = 0, \;\;\;\; [\gamma_5, \hat{\gamma}_{\mu}] = 0.
\end{equation}
These rules are easy to track in standard algebraic manipulation programs.  We use {\tt Tracer}~\cite{Jamin:1991dp} to implement these rules, together with in-house routines written in {\tt Form}~\cite{Kuipers:2012rf} as a cross-check.  

The use of the HV scheme necessitates an additional transformation in order to obtain the standard $\overline{\text{MS}}$ factorization scheme for the PDFs, as is well known in the literature~\cite{Gordon:1993qc,Vogelsang:1995vh,Vogelsang:1996im,Vogt:2008yw,Moch:2008fj,Moch:2014sna}.  Switching to a matrix notation in parton flavors, the beam function computing using Eq.~(\ref{eq:HVBM}) can be written as
\begin{equation}
\Delta B = \Delta \tilde I \otimes \Delta \tilde f,
\end{equation}
where the tilde represents the results before scheme transformation.  We decompose the matching coefficients into singlet and non-singlet pieces in analogy to the usual decomposition performed for the splitting functions:
\bea
\label{eq:flavorsplit}
\Delta {\cal I}_{q_iq_j}(z) =& \Delta {\cal I}_{\bar q_i \bar q_j}(z) = \delta_{ij} \Delta {\cal I}_{qq}^{(V)}(z) 
+\Delta {\cal I}_{qq}^{(S)}(z), \\
\Delta {\cal I}_{q_i \bar q_j}(z) =& \Delta {\cal I}_{\bar q_i q_j}(z) = \delta_{ij} \Delta {\cal I}_{q\bar q}^{(V)}(z) 
+\Delta {\cal I}^{(S)}_{qq}(z).\eea
The scheme transformation can be derived by demanding helicity conservation for massless quarks, which relates the polarized and unpolarized splitting functions.  These requirements naturally extend to the matching coefficients, leading to the relations
\begin{equation}
\Delta I_{qq}^{(V)} = I_{qq}^{(V)},~~~~~~ \Delta I_{q\bar q}^{(V)} = -I_{q \bar q}^{(V)}.
\label{eq-consI}
\end{equation}
The naively-computed $\Delta \tilde I$ do not satisfy these constraints.  We can restore these relations by the transformations
\bea
\Delta B = \left( \Delta \tilde I \otimes \bar Z^5\right) \otimes \left(Z^5 \otimes \Delta \tilde f\right)
= \Delta I \otimes \Delta f
\eea
where 
\bea
\label{eq:HVtrans}
\Delta I = \Delta \tilde I \otimes \bar Z^5, ~~~~
\Delta f = Z^5 \otimes \Delta \tilde f
\eea
with $Z^5$ the scheme transformation matrix, and $\bar Z^5$ is its inverse which satisfies $\bar Z^5\otimes Z^5 = 1$.  More details on the transformation matrix are given later in this manuscript.
%


Having established the factorization scheme transformation and the expansion of the renormalization condition through the necessary NNLO order, we now consider the matching condition.  To obtain the matching coefficients we replace the proton states in Eq.~(\ref{eq:opdef}) by perturbative quark or gluon states.  With this replacement the polarized PDFs can be calculated as an expansion in $\alpha_s$, taking on the familiar forms
\bea
\label{eq:PDFdef}
\Delta \tilde f_{ij}^{(1)} (z) =&  - \frac{1}{\epsilon} \Delta \tilde P_{ij}^{(0)}(z),\nonumber \\
\Delta \tilde f_{ij}^{(2)} (z) =& \frac{1}{2\epsilon^2}\sum_k \Delta \tilde  P_{ik}^{(0)}(z)\otimes \Delta \tilde P_{kj}^{(0)}(z) 
+\frac{\beta_0}{4\epsilon^2}\Delta \tilde P_{ij}^{(0)}(z) - \frac{1}{2\epsilon}\Delta \tilde P_{ij}^{(1)}(z),
\eea
where $\beta_0$ is the usual leading-order QCD beta function,
\begin{equation}
\beta_0 = \frac{11 C_A - 4 T_R N_F}{3}.
\end{equation}
The polarized splitting functions $\Delta \tilde P_{ij}$ needed in our calculation are defined in the Appendix.  
Expanding the matching equation in terms of $\alpha_s$, we can derive the NLO matching coefficient in terms of the renormalized beam function and PDFs:
\bea
\Delta \tilde {\mathcal I}_{ij}^{(1)}(t,z,\mu) = \Delta B_{ij}^{(1)}(t,z,\mu) -2\delta(t)\Delta \tilde f_{ij}^{(1)}\left(z\right).
\label{eq-I1}
\eea
Similarly, we can derive the matching coefficient at order $\alpha_s^2$
\bea
\Delta \tilde {\mathcal I}_{ij}^{(2)}(t,z,\mu) = \Delta B_{ij}^{(2)}(t,z,\mu) - 4\delta(t)\Delta \tilde f_{ij}^{(2)}\left(z\right) 
- 2\sum_k \Delta \tilde {\mathcal I}_{ik}^{(1)}(t,z,\mu) \otimes  \Delta \tilde f_{kj}^{(1)}\left(z\right) \, .
\label{eq-I2nnlo}
\eea
The convolutions required for the calculation were computed with the mathematica package {\tt MT}~\cite{Hoeschele:2013gga}.  Notice that the PDFs of Eq.~(\ref{eq:PDFdef}) and the matching coefficients of Eqs. (\ref{eq-I1}, \ref{eq-I2nnlo}) in our matching calculations still require the scheme transformation of Eq.~(\ref{eq:HVtrans}).  Performing this transformation, we obtain the physical results which restore the helicity conservation equations as shown in Eq. (\ref{eq-consI}).  

As discussed in Ref.~\cite{Stewart:2010qs} and reviewed above, the beam function satisfies a renormalization group equation that allows the logarithmic dependence of the matching coefficients on the renormalization scale $\mu$ to be derived.  Solving this equation allows us to predict all logarithmically-enhanced terms in $t$ in terms of known anomalous dimensions:
\bea
\Delta{\mathcal I}_{ij}^{(1)}(t,z,\mu) =& ~\frac{1}{\mu^2}{\mathcal L}_1\left(\frac{t}{\mu^2}\right) \Gamma_0^i \delta_{ij}\delta(1-z)
+\frac{1}{\mu^2}{\mathcal L}_0\left(\frac{t}{\mu^2}\right)\left[- \frac{\gamma_{B0}^i }{2}\delta_{ij}\delta(1-z) +2\Delta P_{ij}^{(0)}(z)\right]
+\delta(t) \Delta {I}^{(1)}_{ij}(z),
\label{eq-Inlogen}
\\
\Delta{\mathcal I}_{ij}^{(2)}(t,z,\mu) =& ~\frac{1}{\mu^2}{\mathcal L}_3\left(\frac{t}{\mu^2}\right) \frac{(\Gamma_0^i)^2}{2}\delta_{ij}\delta(1-z)
+\frac{1}{\mu^2}{\mathcal L}_2\left(\frac{t}{\mu^2}\right)\Gamma_0^i \left[-\left(\frac{3}{4}\gamma_{B0}^i+\frac{\beta_0}{2}\right)\delta_{ij}\delta(1-z) +3\Delta P_{ij}^{(0)}(z)\right]
\nnu
&+\frac{1}{\mu^2}{\mathcal L}_1\left(\frac{t}{\mu^2}\right) \Bigg\{
\left[\Gamma_1^{i}-(\Gamma_0^i)^2\frac{\pi^2}{6} + \frac{(\gamma_{B0}^i)^2}{4}+\frac{\beta_0}{2}\gamma_{B0}^i\right]\delta_{ij}\delta(1-z)
+\Gamma_0^i\Delta I_{ij}^{(1)}(z) 
\nnu
& - 2(\gamma_{B0}^i+\beta_0)\Delta P_{ij}^{(0)}(z) + 4\sum_k \Delta P_{ik}^{(0)}(z)\otimes \Delta P_{kj}^{(0)}(z)\Bigg\}
\nnu
&+\frac{1}{\mu^2}{\mathcal L}_0\left(\frac{t}{\mu^2}\right) \Bigg\{
\left[(\Gamma_0^i)^2 \zeta_3 + \Gamma_0^i\gamma_{B0}^i\frac{\pi^2}{12}-\frac{\gamma_{B1}^i}{2}\right]\delta_{ij}\delta(1-z)
-\Gamma_0^i\frac{\pi^2}{3}\Delta P_{ij}^{(0)}(z) - \left(\frac{\gamma_{B0}^i}{2}+\beta_0\right)\Delta { I}_{ij}^{(1)}(z)
\nnu
&+ 2\sum_k \Delta {I}^{(1)}_{ik}(z)\otimes \Delta P_{kj}^{(0)}(z) +4 \Delta P_{ij}^{(1)}(z)\Bigg\}
\nnu
&+ \delta(t) \Delta{I}^{(2)}_{ij}(z).
\label{eq-Innlogen}
\eea
The non-cusp anomalous dimension for the quark beam function in $\overline {\rm MS}$ is the same as that in unpolarized case, and reads
\bea
\gamma_{B0}^{q} = & 6C_F, \\
\gamma_{B1}^{q} = & C_F\left[C_A\left(\frac{146}{9}-80\zeta_3\right)+C_F(3-4\pi^2+48\zeta_3)+\beta_0\left(\frac{121}{9}+\frac{2\pi^2}{3}\right)\right].
\eea
The cusp anomalous dimension is
\bea
\Gamma_0^q = 4C_F,~~~~ \Gamma_1^q = C_F\frac{4}{3}\left[(4-\pi^2)C_A+5\beta_0\right].
\eea
 $\mathcal L_n$ is the standard plus distribution, defined as
\bea
\mathcal L_n(x) =\left[\frac{\theta(x)\ln^n(x)}{x}\right]_+ .
\eea
The only terms to determine are the coefficients of the scale-independent $\delta(t)$ contributions, which we label as $\Delta {I}^{(1)}_{ij}(z)$ and  $\Delta{I}^{(2)}_{ij}(z)$.  In our calculation we derive all the scale-dependent terms as well in order to check that our results satisfy Eqs.~(\ref{eq-Inlogen}) and~(\ref{eq-Innlogen}).

\section{Calculation at next-to-leading order}
\label{sec:nlo}

We begin by discussing the calculation of the matching coefficients at NLO.  At this order, there are two matching coefficients to consider, ${\cal I}^{(1)}_{qq}$ and ${\cal I}^{(1)}_{qg}$.  The first can be obtained by setting the external proton to a quark, while the second can be obtained by setting the external proton to a gluon.  We note that ${\cal I}^{(1)}_{\bar{q}\bar{q}}={\cal I}^{(1)}_{qq}$ and ${\cal I}^{(1)}_{\bar{q}g}={\cal I}^{(1)}_{qg}$, while ${\cal I}^{(1)}_{q\bar{q}}=0$.  To perform our calculation we work in light-cone gauge with $n \cdot A=0$.  This has the effect of setting all diagrams with gluons emitted from Wilson lines to zero.  In this gauge there is a single diagram contributing to each channel.  These are shown in Fig.~\ref{fig-nlo}. The cut particles in these diagrams are those satisfying the on-shell constraints.
\bef
\psfig{file=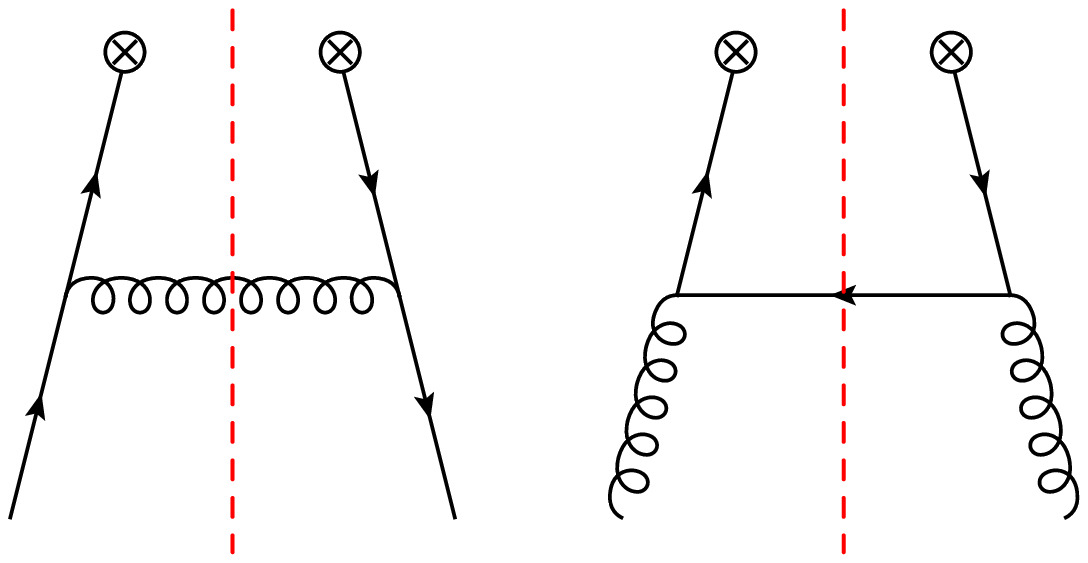, width=3.5in}
\caption{One-loop Feynman diagrams for the quark beam function.  The left diagram determines ${\cal I}^{(1)}_{qq}$, while the right diagram leads to ${\cal I}^{(1)}_{qg}$.  The crosses denote the Wilson lines.  The partons entering these crosses are those which enter the hard scattering process. The red dashed lines denote which intermediate particles are put on-shell.}
\label{fig-nlo}
\eef

We use the quark channel as an example to show the details of our calculation. In light-cone gauge, only the left diagram of Fig.~\ref{fig-nlo} contributes to the quark beam function. This diagram can be calculated by using the standard QCD Feynman rules,
\bea \label{eq:NLOtrace}
\left(\frac{\alpha_s}{4\pi}\right)\Delta B_{qq}^{bare(1)}(t,z)=&\frac{g^2}{N_c}\left(\frac{\mu^2 e^{\gamma_E}}{4\pi}\right)^{\epsilon}
\int d{\rm PS^{(1)}} {\rm Tr}\left[\frac{\slashed{\bar n}\gamma_5}{2}\slashed{\ell}\gamma^{\rho}\mathcal{P}_R\slashed{p}\gamma^{\sigma}\slashed{\ell}\right]d_{\rho\sigma}(k)\frac{1}{\ell^2}\frac{1}{\ell^2}\rm {Tr}[T^aT^a],
\eea
where $\mathcal{P}_R=(1+\gamma_5)/2$ is the spin projection operator.   As mentioned previously we use light-cone gauge for the gluon to avoid extra diagrams with a gluon radiated from gauge link, leading to the propagator numerator
\bea
d_{\rho\sigma}(k)=-g_{\rho\sigma}+\frac{\bar n_{\rho}k_{\sigma}+\bar n_{\sigma}k_{\rho}}{\bar n\cdot k}.
\eea
The final-state phase space at NLO can be written as
\bea
\int d{\rm PS^{(1)}} &= \int \frac{d^dk  }{(2\pi)^{d-1}} d^d\ell \, \, \delta(k^2) \delta(\omega-\ell^-)
\delta(t-\omega k^+)\delta^d(p-k-\ell)\nnu
&= \frac{1}{(4\pi)^{2-\epsilon}}\frac{1}{\Gamma(-\epsilon)}\frac{1}{\omega}\int_0^{t\frac{1-z}{z}} d\hat k_{\perp}^2(\hat k_{\perp}^2)^{-1-\epsilon}.
\label{eq-ps1}
\eea
To obtain the expression in the second line we have separated the final-state phase space into a 4-dimensional part and an $\epsilon$-dimensional part, where $\hat k_{\perp}$ denotes the $\epsilon$-dimensional momentum component.  We have also used the delta functions to simplify the expression.

At this point the calculation of the trace in Eq.~(\ref{eq:NLOtrace}) and the integration over the phase space in Eq.~(\ref{eq-ps1}) are straightforward.  We derive the following bare NLO beam functions:
\bea
\Delta B_{qq}^{bare(1)}(t,z) =&2 \frac{(\mu^2 e^{\gamma_E})^{\epsilon}}{\Gamma(1-\epsilon)} C_F
t^{-1-\epsilon}\bigg(\frac{z}{1-z}\bigg)^{\epsilon}\bigg[\frac{1+z^2}{1-z}+\frac{3+\epsilon}{1-\epsilon}\epsilon(1-z)\bigg],\\
\Delta B_{qg}^{bare(1)}(t,z) =& 2\frac{(\mu^2 e^{\gamma_E})^{\epsilon}}{\Gamma(1-\epsilon)} T_R
t^{-1-\epsilon}\bigg(\frac{z}{1-z}\bigg)^{\epsilon}\bigg[-1+2z-\frac{2\epsilon}{1-\epsilon}(1-z)\bigg].
\eea
Expanding this result to $\mathcal{O}(\epsilon^0)$ (we note that higher-order terms in $\epsilon$ will be needed when using these expressions in the NNLO calculation), we obtain 
\bea
\Delta B_{qq}^{bare (1)}(t,z) =& \frac{4}{\epsilon^2}C_F\delta(t)\delta(1-z)
- \frac{4}{\epsilon}C_F\frac{1}{\mu^2}\mathcal{L}_0\left(\frac{t}{\mu^2}\right)\delta(1-z)
- \frac{2}{\epsilon}C_F\delta(t)\mathcal{L}_0(1-z)(1+z^2)\nnu
&+4C_F\frac{1}{\mu^2}\mathcal{L}_1\left(\frac{t}{\mu^2}\right)\delta(1-z)
+2C_F\frac{1}{\mu^2}\mathcal{L}_0\left(\frac{t}{\mu^2}\right)\mathcal{L}_0(1-z)(1+z^2)\nnu
&+2C_F\delta(t)\bigg[\mathcal{L}_1(1-z)(1+z^2)-\frac{1+z^2}{1-z}\ln z-3(1-z)
-\frac{\pi^2}{6}\delta(1-z)\bigg],
\label{eq-Bqq1bare}
\\
\Delta B_{qg}^{bare (1)}(t,z) =& -\frac{2}{\epsilon}T_R \delta(t)(2z-1)+
2T_R\frac{1}{\mu^2}\mathcal{L}_0\left(\frac{t}{\mu^2}\right) (2z-1)
+2T_R\delta(t)\left[(2z-1)\ln\frac{1-z}{z}+2(1-z)\right].
\eea

Upon rewriting the coefficient of the $1/\epsilon$ pole in terms of the polarized splitting function $\Delta P_{qq}$ that appears in the matching, we can read off the renormalization constant from Eq. (\ref{eq-Bqq1bare}), which is the same as that in the unpolarized case: 
\bea
Z_q^{(1)} = \frac{4}{\epsilon^2}C_F\delta(t)
- \frac{4}{\epsilon}C_F\frac{1}{\mu^2}\mathcal{L}_0\left(\frac{t}{\mu^2}\right)
+\frac{3}{\epsilon}C_F\delta(t).
\eea
We note that there is an overall sign difference in our result for $Z_q^{(1)}$ as compared to other results in the literature~\cite{Ritzmann:2014mka}.  This is simply due to a different relation used here between the bare beam function and the renormalized one.  We believe that the choice we make here is the more conventional one.  Substituting the renormalization constant into~Eq. (\ref{eq-z1}), we can derive the renormalized quark beam function:
\bea
\Delta B_{qq}^{(1)}(t,z,\mu^2) =& - \frac{2}{\epsilon}\delta(t){\Delta P_{qq}^{(0)}(z)}
+4C_F\frac{1}{\mu^2}\mathcal{L}_1\left(\frac{t}{\mu^2}\right)\delta(1-z)
+2C_F\frac{1}{\mu^2}\mathcal{L}_0\left(\frac{t}{\mu^2}\right)\mathcal{L}_0(1-z)(1+z^2)\nnu
&+2C_F\delta(t)\bigg[\mathcal{L}_1(1-z)(1+z^2)-\frac{1+z^2}{1-z}\ln z-3(1-z)
-\frac{\pi^2}{6}\delta(1-z)\bigg],
\label{eq}
\\
\Delta B_{qg}^{(1)}(t,z,\mu^2) =& -\frac{2}{\epsilon} \delta(t)\Delta P_{qg}^{(0)}(z)
+2T_R \frac{1}{\mu^2}\mathcal{L}_0\left(\frac{t}{\mu^2}\right) (2z-1)
+ 2T_R \delta(t)\left[(2z-1)\ln\frac{1-z}{z}+2(1-z)\right],
\eea
where the polarized splitting functions can be found in the Appendix. The last step is to match to the polarized PDFs using Eqs.~(\ref{eq-I1}) and~(\ref{eq:PDFdef}).  We note that the residual $1/\epsilon$ poles cancel, and we reproduce the logarithmic terms in Eq.~(\ref{eq-Inlogen}) upon using the anomalous dimensions presented in Section~\ref{sec:renmatch} and the polarized splitting functions in the Appendix.  This serves as a check on our calculation.

At this point we must match to PDFs and perform the scheme transformation to obtain standard $\overline{\text{MS}}$-scheme expressions.  Referring to the Appendix, the necessary transformation is
\begin{eqnarray}
\Delta I_{qq}^{(1)} &=& \Delta \tilde I_{qq}^{(1)} - z_{qq}^{(1)},\nonumber \\
\Delta I_{qg}^{(1)} &=& \Delta \tilde I_{qg}^{(1)} .
\end{eqnarray}
We can then derive the unknown coefficients of the scale-independent $\delta(t)$ structures in the matching coefficients:
\bea
\Delta {I}_{qq}^{(1)}(z) = &2C_F\bigg[\mathcal{L}_1(1-z)(1+z^2)-\frac{1+z^2}{1-z}\ln z-3(1-z)
-\frac{\pi^2}{6}\delta(1-z)\bigg] + 8C_F (1-z),
\label{eq-Iqqz1}\\
\Delta {I}_{qg}^{(1)}(z) =& 2T_R\left[(2z-1)\ln\frac{1-z}{z}+2(1-z)\right].
\label{eq-Iqgz1}
\eea
We note that the last term in Eq. (\ref{eq-Iqqz1}) arises from the $z_{qq}^{(1)}$ correction to the violation of helicity conservation when using the HV scheme to deal with $\gamma_5$.  We note that these matching coefficients are in agreement with a recent NLO calculation of the matching coefficients for transverse-momentum dependent PDFs~\cite{Gutierrez-Reyes:2017glx}.

\section{Calculation at next-to-next-to-leading order}
\label{sec:nnlo}

Having established the matching coefficients at NLO we move onto the NNLO calculation.  We begin by decomposing the $\delta(t)$ contributions to the matching coefficients into singlet and non-singlet pieces, in analogy to Eq.~(\ref{eq:flavorsplit}):
\bea
\Delta {I}_{q_iq_j}^{(2)}(z) =& \Delta {I}_{\bar q_i \bar q_j}^{(2)}(z) = \delta_{ij} \Delta {I}_{qq}^{(2,V)}(z) 
+\Delta {I}_{qq}^{(2,S)}(z), \\
\Delta {I}_{q_i \bar q_j}^{(2)}(z) =& \Delta {I}_{\bar q_i q_j}^{(2)}(z) = \delta_{ij} \Delta {I}_{q\bar q}^{(2,V)}(z) 
+\Delta {I}_{qq}^{(2,S)}(z),\\
\Delta {I}_{qg}^{(2)}(z) =& \Delta {I}_{\bar qg}^{(2)}(z) = \Delta {I}_{qg}^{(2)}(z).
\eea
As we will see in our explicit calculations, $\Delta {I}_{qq}^{(2,S)}(z)$ is determined by the $q'\to q$ channel, $\Delta {I}_{q\bar q}^{(2,V)}(z)$ is determined in the $\bar q\to q$ channel, $\Delta {I}_{qq}^{(2,V)}(z)$ is determined in the $q\to q$ channel, and $\Delta {I}_{qg}^{(2)}(z)$ is determined in the $g\to q$ channel.  For completeness, we reproduce the necessary scheme transformations of these quantities from the Appendix:
\bea
\Delta I_{qq}^{(2,V)} =& \Delta \tilde I_{qq}^{(2,V)}  - \Delta \tilde I_{qq}^{(1)} \otimes z_{qq}^{(1)} + z_{qq}^{(1)}\otimes z_{qq}^{(1)}
- z_{qq}^{(2,V)},
\nnu
\Delta I_{q\bar q}^{(2,V)} =& \Delta \tilde I_{q\bar q}^{(2,V)} - z_{q\bar q}^{(2,V)},
\nnu
\Delta I_{q q}^{(2,S)} =& \Delta \tilde I_{q q}^{(2,S)} - z_{q q}^{(2,S)}.
\label{eq-IZ5renorm}
\eea

We organize our calculation in terms of cut diagrams, which distinguish whether the two additional partons that appear in the NNLO calculation are virtual or real.  The double-virtual corrections, in which both additional partons are virtual, are scaleless and vanish in dimensional regularization.  This leaves us with real-real and real-virtual diagrams to calculate.  
Fig.~\ref{fig-nnlo_real} shows the symmetric diagrams contributing to the real-real corrections at NNLO as a representative example of the types of contributions which occur.  Interference diagrams are not explicitly shown.  Fig.~\ref{fig-nnlo_virtual} shows the real-virtual corrections at NNLO.  Mirror diagrams are not explicitly shown.

\bef
\psfig{file=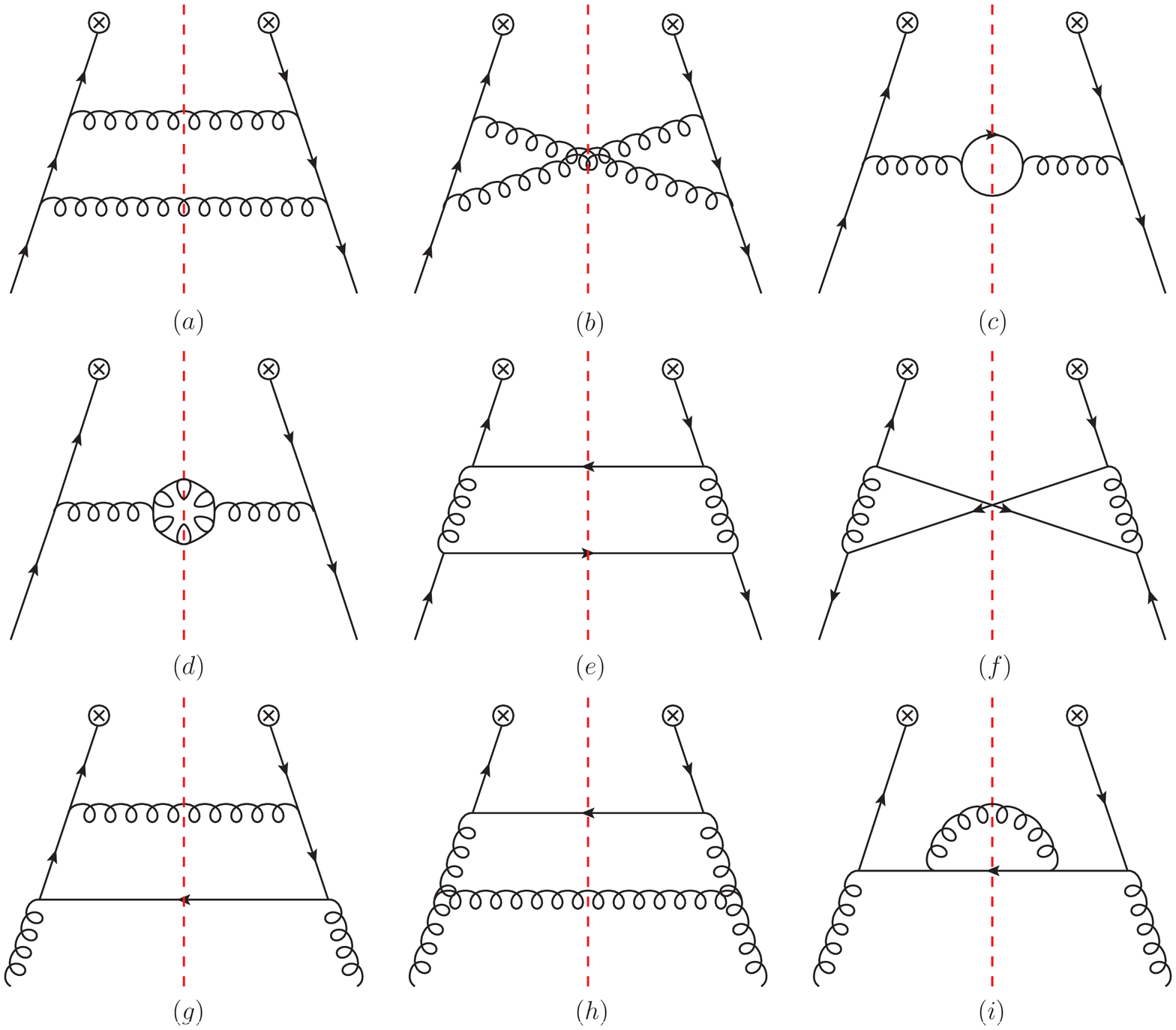, width=5.5in}
\caption{Symmetric real-real diagrams at NNLO.  Interference diagrams are not shown.  Dashed lines indicate which intermediate particles are on-shell.}
\label{fig-nnlo_real}
\eef

\bef
\psfig{file=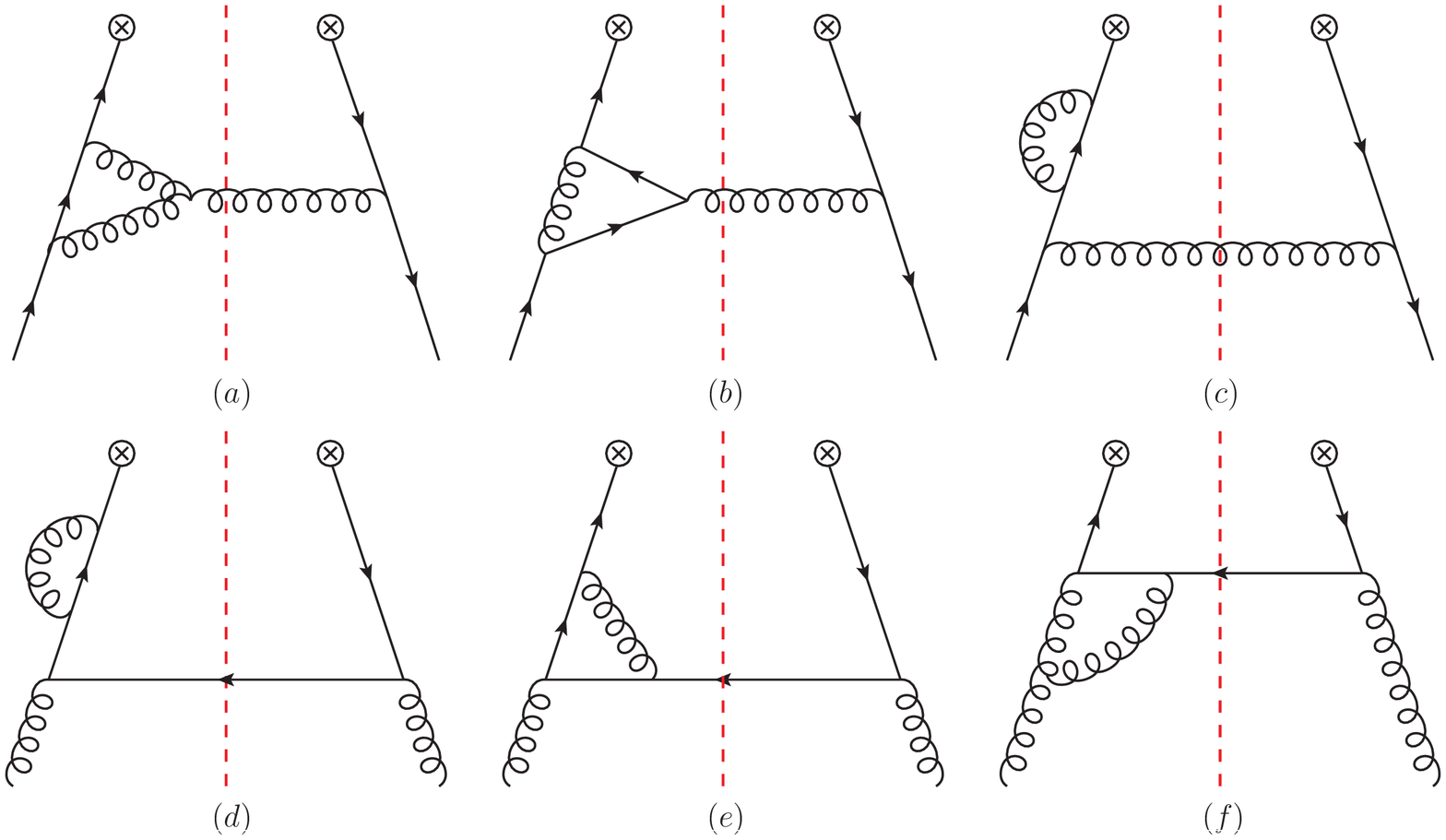, width=5.5in}
\caption{Representative real-virtual diagrams at NNLO.  Mirror diagrams are not explicitly displayed.  Dashed lines indicate which intermediate particles are on-shell.}
\label{fig-nnlo_virtual}
\eef

\subsection{The $q' \to q$  and $\bar{q} \to q$ channels}

We will begin by showing the details of our treatment of the $q' \to q$ transition, since this calculation is sufficiently compact to demonstrate explicitly.  In light-cone gauge there is only one diagram that contributes: Fig.~\ref{fig-nnlo_real}(e), in which the quark entering the hard-scattering cross section is a different flavor than the one coming from the PDF. This diagram can be calculated by using the standard QCD Feynman rules:
\bea
\label{eq:qqptrace}
\left(\frac{\alpha_s}{4\pi}\right)^2\Delta B_{qq'}^{bare(2)}(t,z)=&\frac{g^4}{N_c}\left(\frac{\mu^2 e^{\gamma_E}}{4\pi}\right)^{2\epsilon}
\int d{\rm PS^{(2)}} {\rm Tr}[\mathcal P_R\slashed{p}\gamma^{\nu}\slashed{k_1}\gamma^{\sigma}]
{\rm Tr}[\frac{\slashed{\bar n}\gamma_5}{2}\slashed{\ell}\gamma^{\rho}\slashed{k_2}\gamma^{\mu}\slashed{\ell}]
\frac{1}{(p-k_1)^2}\frac{1}{(p-k_1)^2}\frac{1}{\ell^2}\frac{1}{\ell^2} \nnu
&\times d_{\mu\nu}(p-k_1) d_{\rho\sigma}(p-k_1) \rm {Tr}[T^aT^b]\rm {Tr}[T^aT^b].
\eea
Here, $k_1$ and $k_2$ are the momenta of the intermediate particles that pass the cut.  The final-state phase space for the NNLO real-real correction can be parameterized as
\bea
\int d{\rm PS^{(2)}} &= \int \frac{d^dk_1}{(2\pi)^{d-1}}\frac{d^dk_2}{(2\pi)^{d-1}}d^d\ell \, \, \delta(k_1^2)\delta(k_2^2)\delta(\omega-\ell^-)
\delta\left[t-\omega(k_1^++k_2^+)\right]\delta^d(p-k_1-k_2-\ell).
\label{eq-ps}
\eea
It is straightforward to evaluate the trace appearing in Eq.~(\ref{eq:qqptrace}) in the HV scheme.  We are left with integrals containing the momenta $k_1$ and $k_2$ over the phase space of Eq.~(\ref{eq-ps}).  

We facilitate our calculation using integration-by-parts identities (IBP)~\cite{Tkachov:1981wb,Chetyrkin:1981qh}, implemented in the computer code LiteRed~\cite{Lee:2013mka}.  The only non-standard aspect of our implementation of the IBP identities is our treatment of $\epsilon$-dimensional momenta.  The HV scheme relations of Eq.~(\ref{eq:HVBM}) require us to separate the momenta $k_1$ and $k_2$ into 4-dimensional and $\epsilon$-dimensional pieces, $k_{\mu} = \tilde{k}_{\mu}+\hat{k}_{\mu}$.  Upon doing so we obtain integrals that depend explicitly upon the $\epsilon$-d momenta $\hat{k}_{\mu}$.  We introduce auxiliary vectors that parameterize the $\epsilon$-dimensional direction to handle such contributions.  As an example, suppose our evaluation of the trace in Eq.~(\ref{eq:qqptrace}) contains the dot product $\hat{k}_1 \cdot \hat{k}_2$, and we wish to evaluate the corresponding integral $I^d[\hat{k}_1 \cdot \hat{k}_2]$
that occurs upon integrating this expression over phase space.  We note that $I^d$ can depend upon any other manifestly $d$-dimensional dot products in addition to its dependence on $\hat{k}_1 \cdot \hat{k}_2$.  A simple form-factor decomposition of this integral reveals that we can write
\begin{equation}
I^d[\hat{k}_1 \cdot \hat{k}_2] = -\frac{2\epsilon}{v_{\perp}^2} I^d[(k_1 \cdot v_{\perp}) (k_2 \cdot v_{\perp}))], 
\end{equation}
where $v_{\perp}$ is a space-like vector with support only along the $\epsilon$-dimensional direction.  The integral on the right-hand side is now written in a manifestly $d$-dimensional form, and can be handled using the standard IBP machinery.  Similar relations can be derived for all structures appearing in our integrand.  To obtain all the integrals needed in our calculation we must introduce two such auxiliary momenta, both with support only in the $\epsilon$-dimensional momenta but with an angular separation in this subspace.

After IBP reduction, $\Delta B_{qq'}$ can be expressed in terms of four master integrals: 
\bea
\Delta B_{qq'}^{bare(2)}(t,z)=& C_F T_R \left(\frac{\mu^2 e^{\gamma_E}}{4\pi}\right)^{2\epsilon}\sum_{i=1}^4 C_i(t,z,\epsilon) I_i^{RR} .
\label{IBP-res}
\eea
The four needed master integrals can be derived through direct phase space integration:
\bea
I^{RR}_1&=\int d{\rm PS^{(2)}} \times 1 = 4 \frac{ (16 \pi)^{-3+2 \epsilon}  }{\omega  \Gamma \left(\frac{3}{2}-\epsilon \right)^2} t^{1-2 \epsilon }  \left(\frac{1-z}{z}\right)^{1-2 \epsilon }  ,
\\
I^{RR}_2&= \int d{\rm PS^{(2)}} \times \frac{1}{\bar n \cdot (p-k_1)}= 4 \frac{ (16 \pi)^{-3+2 \epsilon} }{\omega ^2 \Gamma
   \left(\frac{3}{2}-\epsilon \right)^2} t^{1-2 \epsilon }   (1-z)^{1-2 \epsilon } z^{2\epsilon}  \, _2F_1(1,1-\epsilon ;2-2 \epsilon ;1-z) ,
   \\
I^{RR}_3&=\int d{\rm PS^{(2)}} \times \frac{1}{(p-k_1-k_2)^2} = -4 \frac{(16 \pi)^{-3+2 \epsilon} }{\omega  \Gamma
   \left(\frac{3}{2}-\epsilon \right)^2} t^{-2 \epsilon } (1-z)^{1-2 \epsilon } z^{2\epsilon } \, _2F_1(1,1-\epsilon ;2-2 \epsilon ;1-z) ,
   \\
I^{RR}_4&=   \int d{\rm PS^{(2)}} \times \frac{1}{(p-k_1-k_2)^2}\frac{1}{\bar n \cdot (p-k_1)}
 =-4 \frac{ (16 \pi) ^{-3+2 \epsilon} }{\omega ^2 \Gamma
   \left(\frac{3}{2}-\epsilon \right)^2} t^{-2 \epsilon } (1-z)^{1-2 \epsilon } z^{1+2\epsilon} \, _2F_1(1,1-\epsilon ;2-2 \epsilon ;1-z){}^2 .
\eea
We note that there is no $z\to 1$ divergence in this channel, but there is a $t \to 0$ divergence that occurs upon combining these integrals.  Dimensional analysis tells us that the result of the full calculation must be proportional to $t^{-1-2\epsilon}$.  We expand this factor together with the renormalization scale $\mu$ in terms of plus distributions as follows:
\begin{equation}
\frac{1}{\mu^2} \left( \frac{t}{\mu^2} \right)^{-1-2\epsilon} = -\frac{1}{2\epsilon} \delta(t) + \frac{1}{\mu^2} \sum_{n=0} 
\frac{(-2\epsilon)^n}{n!} \mathcal L_n \left(\frac{t}{\mu^2} \right). 
\end{equation}

Upon performing this expansion we obtain the bare beam function $\Delta B^{bare(2)}_{qq'}$.  It is clear from Eq.~(\ref{eq-z2}) that the renormalized and bare beam functions are the same at this order for this channel.  We can use Eq.~(\ref{eq-I2nnlo}) to express the beam function in terms of the matching coefficient:
\bea
\Delta \tilde {\mathcal I}_{qq'}^{(2)}(t,z,\mu) = \Delta B_{qq'}^{(2)}(t,z,\mu) - \frac{2}{\epsilon^2} \delta(t)\Delta \tilde P_{qg}^{(0)}(z)\otimes \Delta \tilde P_{gq'}^{(0)}(z) 
+\frac{2}{\epsilon}\delta(t)\Delta \tilde P_{qq'}^{(1)}(z)
+ \frac{2}{\epsilon} \Delta \tilde {\mathcal I}_{qg}^{(1)}(t,z,\mu) \otimes \Delta \tilde P_{gq'}^{(0)}\left(z\right).
\label{eq-Iqqp2match}
\eea
We note that the one-loop matching coefficient $\Delta \tilde {\mathcal I}_{qg}^{(1)}$ must be derived to higher orders in $\epsilon$ when used in this relation.  All poles cancel in this expression for $\Delta \tilde {\mathcal I}_{qq'}^{(2)}$, which is a powerful check of our calculation.  We are left with the finite matching coefficient, which reads as follows after scheme transformation:
\bea
\Delta{\mathcal I}_{qq'}^{(2)}(t,z,\mu) =& \frac{1}{\mu^2}\mathcal L_1\left(\frac{t}{\mu^2}\right) 4\Delta P_{qg}^{(0)}\otimes \Delta P_{gq'}^{(0)}
+ \frac{1}{\mu^2}\mathcal L_0 \left(\frac{t}{\mu^2}\right) \left[4\Delta P_{qq'}^{(1)} + 2 \Delta I_{qg}^{(1)}(z) \otimes \Delta P_{gq'}^{(0)}\right]\nnu
&+ \delta(t)C_F T_R\Bigg\{ 
(1+z)\bigg[ -8\text{Li}_3(1-z)+8\ln (z)\text{Li}_2(z)-8\ln (1-z)\text{Li}_2(z)+\frac{10}{3} \ln ^3(z)-4\ln(z)\ln^2(1-z)
\nnu
& -\frac{8}{3} \pi ^2\ln (z)+\frac{4}{3}\pi ^2 \ln (1-z) \bigg]
+(1-z)\bigg[4\text{Li}_2(z)-4\pi^2+\frac{13-17 z}{1-z} \ln^2(z)+ 10\ln^2(1-z)
\nnu
&-20\ln (z)\ln(1-z)\bigg] + \left(38-2z\right) \ln (z) - 24(1-z)\ln (1-z) + 64(1-z)
 \Bigg\}.
\eea
Comparing to Eq.~(\ref{eq-Innlogen}) we can confirm that the plus distributions in $t$ have the required coefficients, which serves as a second check of our result.  We isolate the desired scale-independent matching coefficient: 
\bea
\Delta{I}_{qq'}^{(2)}(z) =&\Delta{I}_{qq'}^{(2,S)}(z)
\nnu
=& C_F T_R\Bigg\{ 
(1+z)\bigg[ -8\text{Li}_3(1-z)+8\ln (z)\text{Li}_2(z)-8\ln (1-z)\text{Li}_2(z)+\frac{10}{3} \ln ^3(z)-4\ln(z)\ln^2(1-z)
\nnu
& -\frac{8}{3} \pi ^2\ln (z)+\frac{4}{3}\pi ^2 \ln (1-z) \bigg]
+(1-z)\bigg[4\text{Li}_2(z)-4\pi^2+\frac{13-17 z}{1-z} \ln^2(z)+ 10\ln^2(1-z)
\nnu
&-20\ln (z)\ln(1-z)\bigg] + \left(38-2z\right) \ln (z) - 24(1-z)\ln (1-z) + 64(1-z)
 \Bigg\}.
\eea
We further verify our identification of this term as $\Delta{I}_{qq'}^{(2,S)}$ in our calculations of the $\bar q\to q$ and $q\to q$ channels.  We note that all master integrals appearing in this channel are well-defined in dimensional regularization, without the need for an additional regulator.  A nearly identical calculation for the $\bar q \to q$ channel gives the result
\bea
\Delta{I}_{q\bar q}^{(2,V)}(z) =& C_F \left(\frac{C_A}{2}-C_F\right)  \Bigg\{  
16\frac{1+z^2}{1+z}\bigg[\text{Li}_3\left(\frac{1-z}{2}\right)-
\text{Li}_3(1-z)-\frac{1}{2} \text{Li}_3(z)+ \text{Li}_3\left(\frac{1-z}{1+z}\right)
\nnu
 &  + \text{Li}_3\left(\frac{1+z}{2}\right) + \ln \left(\frac{z}{1-z}\right)\text{Li}_2\left(\frac{1}{1+z}\right)
 -\frac{1}{8} \ln ^3(z) - \frac{1}{2}\zeta (3)- \frac{1}{6}  \ln ^3(1+z) - \frac{1}{3} \ln ^3(2)
 \nnu
 &+\frac{1}{2}  \ln ^2(1+z)\ln(z)+\frac{1}{2}  \ln^2(2) \ln (1-z^2)- \frac{\pi ^2}{6} \ln (1+z)-\ln (2) \ln (1-z) \ln (1+z)
 \nnu
 &+\frac{\pi ^2}{12}\ln(1-z)+\frac{\pi ^2}{6}\ln(2) \bigg]
 - 8(1+z) \text{Li}_2\left(\frac{1}{1+z}\right) - 8(1+z)\text{Li}_2(z) - 8z \ln ^2(z) 
 \nnu
 &-4 (1+z) \ln ^2(1+z)+2(1+z)\pi^2 + 2(3+19z)\ln(z) + 16(1-z)\ln(1-z) + 30 (1-z) \Bigg\}.
\eea
As a final check of our result, we note that the master integrals we obtain for the polarized beam function are identical to those that appear for the unpolarized one.  It is simple for us to re-evaluate the Feynman diagrams without any polarization in order to obtain the standard unpolarized matching coefficients.  We agree exactly with the results of Ref.~\cite{Gaunt:2014xga} for these channels.  We note that $\Delta{I}_{q\bar q}^{(2,V)}$ satisfies the helicity conservation relation of Eq.~(\ref{eq-consI}).

\subsection{The $g\to q$ and $q \to q$ channels}

We now discuss aspects of the calculation that first appear when we consider $\Delta {I}_{qg}^{(2)}$.  This channel has both real-virtual corrections and a non-trivial renormalization.  The real-real corrections require nine independent master integrals.  We use the method of differential equations~\cite{Kotikov:1990kg} to calculate them, facilitated by the use of the canonical form~\cite{Henn:2013pwa}. This special form of the differential equation was found through the Magnus algorithm~\cite{Argeri:2014qva}.  Since they otherwise present no new feature beyond those already presented in the  $q' \to q$ channel, we do not discuss them further.

After performing the IBP reductions for the real-virtual diagrams we find three master integrals which can be solved by direct integration:
\bea
I^{RV}_1=&\int \frac{d^dk}{(2\pi)^d} 2\pi \delta(k^2) \delta(p^- - k^- - \omega)\delta(t-\omega k^+)
\int\frac{d^dq}{(2\pi)^d}\frac{1}{(p-q)^2(q-k)^2}\\
=&\frac{i}{(4\pi)^{4-2\epsilon}}\frac{t^{-2\epsilon}}{\omega}\frac{\Gamma(1-\epsilon)\Gamma(\epsilon)}{\Gamma(2-2\epsilon)}z^{2\epsilon}(1-z)^{-\epsilon},\\
I^{RV}_2=&\int \frac{d^dk}{(2\pi)^d} 2\pi \delta(k^2) \delta(p^- - k^- - \omega)\delta(t-\omega k^+)
\int\frac{d^dq}{(2\pi)^d}\frac{1}{q^2 (p-q)^2 (q-k)^2 ~ (\bar n \cdot q - \bar n \cdot k) }\\
=&-\frac{i}{(4\pi)^{4-2\epsilon}}\frac{t^{-1-2\epsilon}}{\omega p^-}\frac{1}{\epsilon}
\frac{\Gamma(1+\epsilon)\Gamma^2(-\epsilon)}{\Gamma(1-\epsilon)\Gamma(-2\epsilon)}z^{1+2\epsilon}(1-z)^{-1-\epsilon}
\, _2F_1(1,-\epsilon,1-\epsilon,\frac{1}{1-z}),\\
I^{RV}_3=&\int \frac{d^dk}{(2\pi)^d} 2\pi \delta(k^2) \delta(p^- - k^- - \omega)\delta(t-\omega k^+)
\int\frac{d^dq}{(2\pi)^d}\frac{1}{q^2 (p-q)^2 (q-k)^2 ~ (\bar n \cdot p - \bar n \cdot q) }\\
=&\frac{i}{(4\pi)^{4-2\epsilon}}\frac{t^{-1-2\epsilon}}{\omega p^-}\frac{1}{\epsilon}
\frac{\Gamma(1+\epsilon)\Gamma^2(-\epsilon)}{\Gamma(1-\epsilon)\Gamma(-2\epsilon)}z^{1+2\epsilon}(1-z)^{-\epsilon}
\, _2F_1(1,-\epsilon,1-\epsilon,1-z).
\eea
We have let $q$ be the loop momentum while $k$ is the momentum for the on-shell parton crossing the cut. In all three master integrals, the final-state phase space integration is trivial,
\bea
\int \frac{d^dk}{(2\pi)^d} 2\pi \delta(k^2) \delta(p^- - k^- - \omega)\delta(t-\omega k^+)
=\frac{1}{(4\pi)^{2-\epsilon}}\frac{1}{\Gamma(1-\epsilon)}\frac{1}{\omega}t^{-\epsilon}\left(\frac{z}{1-z}\right)^{\epsilon}.
\eea
This fixes the momentum components of $k$ as: $k^+=\frac{t}{\omega},~k^-=p^- - \omega=(1-z)p^-,~ k_{\perp}^2 = k^+k^-=\frac{1-z}{z}t$. The rest of the integration can be done by using standard Feynman parameterization techniques.

We note that all of these master integrals are well-defined in dimensional regularization, as are all the real-real master integrals.  No new master integral appears in the calculation of the $q \to q$ channel, meaning that our entire calculation is well-defined without the introduction of an additional regulator.  In addition, each cut diagram is well-defined without an additional regulator, since each can be expressed as a linear combination of master integrals with rational coefficients depending on $\epsilon$ and $z$.  The calculation of the unpolarized beam function of Ref.~\cite{Gaunt:2014xga} was stated to require a regulator beyond dimensional regularization in intermediate stages.  Since it is completely straightforward for us to also apply our master integrals to the calculation of the unpolarized beam function, we have done so and found complete agreement with the matching coefficients of Ref.~\cite{Gaunt:2014xga}, while finding that no additional regulator is needed for that calculation either.  

We now detail the matching and renormalization in the $g\to q$ channel, where ultraviolet divergences first appear.
The first step is to take into account the renormalization of the strong coupling constant in the $\overline{\rm MS}$ scheme, for which we need the following renormalization constant:
\bea
Z_{\alpha}=1-\frac{\alpha_s}{4\pi}\frac{\beta_0}{\epsilon}+{\cal O}(\alpha_s^2).
\eea
The two-loop contribution to the bare beam function after $\alpha_s$ renormalization is
\bea
\Delta B_{qg}^{bare(2)}(t,z) = \Delta B_{qg}^{bare(2)}(\alpha_s^{(0)},t,z) - \frac{\beta_0}{\epsilon} \Delta B_{qg}^{bare(1)}(\alpha_s^{(0)},t,z).
\eea
From the relation between the bare and renormalized beam functions we can derive the renormalized beam function in the $g\to q$ channel as
\bea
\Delta B_{qg}^{(2)}(t,z,\mu) = \Delta B_{qg}^{bare(2)}(t,z) - \int dt' Z_{q}^{(1)}(t-t',\mu)  \Delta B_{qg}^{(1)}(t',z,\mu),
\eea
where $Z_{q}^{(1)}$ and $\Delta B_{qg}^{(1)}(t,z,\mu)$ have already been determined in Section~\ref{sec:nlo}.  We note that the one-loop beam function is needed to ${\cal O}(\epsilon^2)$ in this equation.  With the renormalization finished, we next consider the matching.  For the $g\to q$ channel, it is
\bea
\begin{split}
\Delta\tilde {\mathcal I}_{qg}^{(2)}(t,z,\mu) &= \Delta B_{qg}^{(2)}(t,z,\mu) - 4\delta(t)\Delta \tilde f_{qg}^{(2)}\left(z\right) 
- 2 \Delta\tilde {\mathcal I}_{qq}^{(1)}(t,z,\mu) \otimes  \Delta \tilde f_{qg}^{(1)}\left(z \right)- 2 \Delta\tilde {\mathcal I}_{qg}^{(1)}(t,z',\mu) \otimes  \Delta \tilde f_{gg}^{(1)}\left(z \right),
\end{split}
\eea
where the polarized PDFs are defined in Eq. (\ref{eq:PDFdef}).
%
%
We have verified that all the divergences vanish in the matching, as required. After scheme transformation, the logarithmic terms all agree with the constraints shown in Eq. (\ref{eq-Innlogen}), providing a check on our result.  We present here the desired scale-independent coefficient of the $\delta(t)$ term:
\bea
\Delta{I}_{qg}^{(2)}(z) =&
C_FT_R\Bigg\{2(1-2 z) \Big[10\text{Li}_3(z)+ 12\text{Li}_3(1-z) + 4\ln(1-z) \text{Li}_2(z) +\pi^2\ln(1-z) - 3\ln(1-z)\ln^2z  - 4\text{Li}_2(z)\ln(z)
\nnu
&~~  - \frac{7}{3} \pi ^2 \ln (z) -32\zeta (3) - \frac{5}{3} \ln ^3(1-z) + \frac{1}{6} \ln ^3(z) + 8 \ln(z)\ln^2(1-z)\Big] - 6(7-8 z) \ln(z)\ln (1-z)
\nnu
&~~ -2(3-4 z) \text{Li}_2(z) - \frac{2}{3} (7-8 z)\pi ^2 +\bigg(21-24 z\bigg) \ln^2(1-z)+\left(\frac{37}{2}-12 z\right) \ln ^2(z) 
\nnu
&~~-12 (5-6 z)\ln (1-z) +\bigg(70-63z\bigg)\ln (z) + 99-109 z \Bigg\}
\nnu
&+C_AT_R\Bigg\{
2(1+2 z) \bigg[4\text{Li}_3\left(\frac{1-z}{2}\right) + 4\text{Li}_3\left(\frac{1-z}{1+z}\right) + 4\text{Li}_3\left(\frac{1+z}{2}\right)+4\ln\left(\frac{z}{1-z}\right) \text{Li}_2\left(\frac{1}{1+z}\right) 
\nnu
&~~ + 2 \ln\left(\frac{z}{1+z}\right)\ln^2(1+z) + \frac{4}{3}\ln^3\left(\frac{1+z}{2}\right) + 2\ln(2) \ln\left(\frac{1+z}{1-z}\right)\ln \left(\frac{(1+z)^2}{2}\right)-\frac{2}{3}\pi ^2 \ln \left(\frac{1+z}{2}\right)\bigg]
\nnu
&~~  -2(1-2 z) \bigg[4\ln^2(z)\ln(1-z) +\frac{1}{3}\ln^3(1-z)\bigg] - 40 \text{Li}_3(1-z)-16 z \text{Li}_3(z)-4 (5+2 z) \text{Li}_2(z)\ln(1-z)
\nnu
&~~  +\frac{2}{3} \pi ^2 (7+2 z)\ln(1-z) +\frac{2}{3} (7+10 z) \ln ^3(z) + 8(1+4z) \text{Li}_2(z)\ln(z) - \frac{16}{3} (1+z) \ln(z)\pi^2
\nnu
&~~ + 2(3-14 z) \zeta (3) - 8 (1+z) \ln (z)\ln^2(1-z) - 4(1+4z) \text{Li}_2(z)+8 (1+z) \text{Li}_2\left(\frac{1}{1+z}\right)
\nnu
&~~ - 48 (1-z) \ln (z)\ln(1-z) - \frac{2}{3} (11-14 z) \pi ^2 +20 (1-z) \ln^2(1-z) +\left(29-44 z\right)\ln^2(z)
\nnu
&~~  +4 (1+z) \ln^2(1+z) -2(12-13 z)\ln(1-z) + 80\ln(z) +2(57-59 z) 
\Bigg\}.
\eea

The only new feature present in the calculation of the $q \to q$ channel is the appearance of $z \to 1$ singularities.  They are straightforwardly extracted by a standard expansion in plus distributions:
\begin{equation}
(1-z)^{-1+ n\epsilon} = \frac{1}{n\epsilon}  +\sum_{i=0} \frac{(n\epsilon)^i}{i!} \mathcal L_i (1-z).
\end{equation}
We note that the needed renormalization factor turns out to be equivalent to the one found for the unpolarized case~\cite{Ritzmann:2014mka}.  The desired scale-independent matching coefficient is
\bea
\Delta{I}_{qq}^{(2)}(z) = \Delta{I}_{qq}^{(2,V)}(z) + \Delta{I}_{qq}^{(2,S)}(z),
\eea
where
\bea
\Delta{I}_{qq}^{(2,V)}(z) =&\delta(1-z)C_F\bigg[C_F\frac{7\pi^4}{30} +C_A\left(\frac{208}{27}-\frac{2\pi^2}{3}-\frac{\pi^4}{9}\right)
+\beta_0\left(\frac{164}{27}-\frac{5\pi^2}{6}-\frac{10\zeta_3}{3}\right)\bigg]
\nnu
&+C_F^2\Bigg\{4\bigg[8\zeta (3) {\mathcal L}_0(1-z) - \frac{5 \pi ^2 }{3}{\mathcal L}_1(1-z) + 2 {\mathcal L}_3(1-z)\bigg]
+\frac{1}{1-z}\bigg[8 \left(1+2 z^2\right)\text{Li}_3(1-z)
\nnu
&~~  - 28 \left(1+z^2\right) \text{Li}_3(z)  - 8 z^2\ln(1-z)\text{Li}_2(1-z) + 2 \left(z^2-3\right) \ln ^2(z)\ln(1-z) 
\nnu
&~~ + \frac{10}{3} \pi ^2 (1-z^2)\ln(1-z)
+ \frac{ 22+26z^2}{3} \pi ^2 \ln(z) -8 \left(2 z^2+3\right) \text{Li}_2(1-z)\ln(z)  
\nnu
&~~ + 4 \left(3+11 z^2 \right) \zeta (3) - 4 \left(3+4 z^2\right) \ln (z) \ln ^2(1-z)
- 4 (1-z^2) \ln ^3(1-z) +\frac{5}{3} (1-z^2) \ln ^3(z) \bigg]
\nnu
&~~ +\frac{1}{1-z}\bigg[ -4 (4 + z -2z^2) \ln (z)\ln(1-z) -12(1+z-z^2)\text{Li}_2(1-z) + \frac{2}{3} (1-z)^2 \pi ^2  
\nnu
&~~+ \left(7+14z -12 z^2  \right) \ln ^2(z)\bigg]
- 2\left(12 - 11 z\right)\ln (1-z) + 2\left(-5-28 z+\frac{8}{1-z}\right) \ln (z) -26+30 z
\Bigg\}
\nnu 
&+C_FC_A\Bigg\{
  \left(28 \zeta (3)-\frac{64}{9}\right) {\mathcal L}_0(1-z)  -\left(\frac{4 \pi ^2}{3} - \frac{16}{3} \right) {\mathcal L}_1(1-z)
+\frac{1+z^2}{1-z}\bigg[-16\text{Li}_3(1-z) 
\nnu
&~~+4\text{Li}_3(z) +4 \text{Li}_2(1-z)\ln(1-z) + 4 \ln ^2(z)\ln(1-z)+\frac{2}{3} \frac{1-z^2}{1+z^2}\pi^2\ln(1-z)
+ 8\text{Li}_2(1-z)\ln(z) 
\nnu
&~~  - \ln ^3(z)  - \frac{2 \left(9-5 z^2 \right) }{1+z^2}\zeta(3) \bigg] + 4 (1+z) \ln (z)\ln(1-z) + 4 (1+z) \text{Li}_2(1-z) - \frac{1-z}{3} \pi ^2
 \nnu
 &~~ -6z \ln ^2(z) + \frac{1}{3}\left(28-38z\right)\ln (1-z) + \left(3+27z - \frac{10}{1-z}\right) \ln (z)+\frac{1}{9} (95-67 z)
\Bigg\}
\nnu
&+C_F\beta_0\Bigg\{
\left(\frac{2 \pi ^2}{3}-\frac{56}{9}\right) {\mathcal L}_0(1-z)+\frac{20 }{3}{\mathcal L}_1(1-z) -2 {\mathcal L}_2(1-z)
+\frac{1+z^2}{1-z}\bigg[2 \text{Li}_2(1-z) - \frac{5 }{2}\ln ^2(z)
\nnu
&~~+4\ln(z)\ln(1-z) - \frac{1}{3} \frac{1-z^2}{1+z^2}\pi ^2 + \frac{1-z^2}{1+z^2} \ln^2(1-z)\bigg]
-\frac{4}{3} (1+4 z)\ln (1-z)
\nnu
&~~ - \frac{5-2z+7z^2}{1-z} \ln (z)+\frac{1}{9} (37+19 z)\Bigg\}.
\eea
We have applied our techniques to also calculate the unpolarized matching coefficients, finding complete agreement with the results of Ref.~\cite{Gaunt:2014xga}.  We note that $\Delta{I}_{qq}^{(2,V)}$ satisfies the helicity conservation relation of Eq.~(\ref{eq-consI}).

\section{Conclusions}
\label{sec:conc}

In this paper we have calculated the longitudinally-polarized quark beam function through NNLO in QCD perturbation theory.  This is the last remaining quantity needed to apply the $N$-jettiness factorization theorem through the NNLO level to polarized collisions.    We expect that our result for $\Delta B$ will be useful in numerous applications, from enabling NNLO results for polarized collisions at RHIC and future colliders, to the study of event shapes at a future EIC.  We have presented in detail the technical features of our derivation, which we believe will be useful in future calculations of similar quantities.  As a by-product of our calculation we have provided the first independent check of the unpolarized quark beam function obtained in Ref.~\cite{Gaunt:2014xga}.  This quantity has been heavily used in the $N$-jettiness subtraction approach to fixed-order calculations, and an independent derivation of the result was highly desirable.  We have shown that the NNLO beam function can be obtained using only dimensional regularization to handle all singularities.  We look forward to phenomenological applications of our results. 

\section*{Acknowledgements}

We thank A.~Vladimirov for questions regarding the factorization scheme dependence in an initial version of this manuscript, which led to corrections in our NNLO matching coefficients and an expanded discussion of this topic.  We thank X.~Liu and P.~Mastrolia for helpful conversations.  R.~B. and U.~S. are supported by the DOE contract DE-AC02-06CH11357.  F.~P. and H.~X. are supported by the DOE grants DE-FG02-91ER40684 and DE-AC02-06CH11357.  This research was also supported in part by the NSF under Grant No. NSF PHY11-25915 to the Kavli Institute of Theoretical Physics in Santa Barbara, which we thank for hospitality during the completion of this manuscript.

\appendix
\section{Polarized splitting functions}

We reproduce here the polarized splitting functions needed in our calculation for both before and after scheme transformation. These are consistent with the literature~\cite{Vogelsang:1995vh,Vogelsang:1996im}. The leading-order results are
\bea
\label{eq-pqq0}
\Delta P_{qq}^{(0)}(z) =& \Delta \tilde P_{qq}^{(0)}(z) = C_F\bigg[\frac{1+z^2}{(1-z)_+}+\frac{3}{2}\delta(1-z) \bigg],
\\
\Delta P_{gq}^{(0)}(z) =& \Delta \tilde P_{gq}^{(0)}(z) = C_F(2-z) ,\\
\Delta P_{qg}^{(0)}(z) =& \Delta \tilde P_{qg}^{(0)}(z) = -T_R(1-2z) , \\
\Delta P_{gg}^{(0)}(z) =& \Delta \tilde P_{gg}^{(0)}(z) =  2C_A\bigg[\frac{1}{(1-z)_+}+1-2z\bigg]+\frac{\beta_0}{2}\delta(1-z).
\eea
The NLO results are
\bea
\Delta P_{qq}^{(1)}(z) =& \delta_{ij}\Delta P_{qq}^{(1,V)}(z) + \Delta P_{qq}^{(1,S)}(z),\\
\Delta P_{q\bar q}^{(1)}(z) =& \delta_{ij}\Delta P_{q\bar q}^{(1,V)}(z) + \Delta P_{qq}^{(1,S)}(z),\\
\Delta P_{qg}^{(1)}(z) =& \Delta P_{\bar qg}^{(1)}(z),
\eea
with the scheme transformation relations shown below:
\bea
\Delta P_{qq}^{(1,V)}(z) =& \Delta \tilde P_{qq}^{(1,V)}(z) - \frac{1}{4}\beta_0 z_{qq}^{(1)}, \\
\Delta P_{q\bar q}^{(1,V)}(z) =& \Delta \tilde P_{q\bar q}^{(1,V)}(z),\\
\Delta P_{q\bar q}^{(1,S)}(z) =& \Delta \tilde P_{q\bar q}^{(1,S)}(z)\\
\Delta P_{qg}^{(1)}(z) =& \Delta \tilde P_{qg}^{(1)}(z) + \frac{1}{2}z_{qq}^{(1)}\otimes \Delta \tilde P_{qg}^{(0)}.
\eea
The scheme transformation factors are defined in Appendix \ref{app-st}, and the final physical splitting functions are
\bea
\Delta P_{qq}^{(1,S)}(z) =& C_F T_R \bigg[1-z-(1-3z)\ln z - (1+z)\ln^2 z\bigg],\\
\Delta P_{q\bar q}^{(1,V)}(z) 
=& C_F \left(\frac{C_A}{2}-C_F\right) \bigg\{ 2p_{qq}(-z)S_2(z)+2(1+z)\ln z +4(1-z)\bigg\},\\
\Delta P_{qq}^{(1,V)}(z) 
=& C_F^2\bigg\{-\bigg[2\ln z \ln(1-z) +\frac{3}{2}\ln(z)\bigg]p_{qq}(z)
-\left(\frac{3}{2}+\frac{7}{2}\right)\ln z - \frac{1}{2}(1+z)\ln^2z - 5(1-z)
\bigg\} \nnu
&+C_FC_A\bigg[\bigg(\frac{1}{2}\ln^2 z +\frac{11}{6}\ln z+\frac{67}{18}-\frac{\pi^2}{6}\bigg)p_{qq}(z)
+(1+z)\ln z + \frac{20}{3}(1-z)\bigg]\nnu
&+C_FT_R n_f\bigg[-\bigg(\frac{2}{3}\ln z +\frac{10}{9}\bigg)p_{qq}(z)-\frac{4}{3}(1-z)\bigg]\nnu
&+\bigg\{C_F^2\bigg[\frac{3}{8}-\frac{\pi^2}{2}+6\zeta(3)\bigg]
+C_FC_A\bigg[\frac{17}{24}+\frac{11}{18}\pi^2-3\zeta(3)\bigg]
-C_FT_R n_f\bigg(\frac{1}{6}+\frac{2}{9}\pi^2\bigg)\bigg]\bigg\}\delta(1-z),
\\
\Delta P_{qg}^{(1)}(z) = & \frac{C_F T_R}{2}\bigg\{-22+27z-9\ln z +8(1-z)\ln(1-z)+p_{qg}(z)
\bigg[2\ln^2(1-z)-4\ln(1-z)\ln z+\ln^2z-\frac{2}{3}\pi^2\bigg]\bigg\}\nnu
&+\frac{C_AT_R}{2}\bigg\{2(12-11z)-8(1-z)\ln(1-z) + 2(1+8z)\ln z 
-2\bigg[\ln^2(1-z)-\frac{\pi^2}{6}\bigg]p_{qg}(z)\nnu
&-\big[S_2(z)-3\ln^2z\big]p_{qg}(-z)\bigg\},\\
\eea
with
\bea
p_{qq}(z) =& \frac{2}{1-z} - 1-z,\\
p_{qg}(z) =& -1+2z,\\
S_2(z) =& \int_{\frac{z}{1+z}}^{\frac{1}{1+z}}\frac{dx}{x}\ln\frac{1-x}{x}
=\frac{1}{2}\ln^2(z)-\ln(z)\ln(1+z) - {\rm Li}_2\left(\frac{1}{1+z}\right) + {\rm Li}_2\left(\frac{z}{1+z}\right).
\eea
We note that contrary to previous definitions of these splitting functions~\cite{Vogelsang:1995vh,Vogelsang:1996im} we have not included a factor of $2 N_F$ in $\Delta P_{qg}^{(0)}$, $\Delta P_{qg}^{(1)}$ and $\Delta P_{qq}^{S(1)}$. 

\section{Factorization scheme transformation}
\label{app-st}
In our calculation, we consider the following scheme transformation matrix
\bea
Z^5 =& 1 + \sum_{n=1} a_s^n Z^{5(n)}\nnu
=& 1+ \sum_{n=1} a_s^n \left(\begin{array}{cc} z_{qq}^{(n)} & z_{qg}^{(n)}\\
z_{gq}^{(n)} & z_{gg}^{(n)} \end{array}\right)\nnu
=& 1 + a_s \left(\begin{array}{cc} z_{qq}^{(1)} & 0\\
0 & 0 \end{array}\right)
+ a_s^2 \left(\begin{array}{cc} z_{qq}^{(2)} & 0\\
0 & 0 \end{array}\right) + \dots
\eea
where $a_s = \alpha_s/(4\pi)$ and $z_{qq} = z_{qq}^V + z_{q\bar q}^V + z_{qq}^S$. In the above equation, we have dropped $z_{qg}^{(n)}$, $z_{gg}^{(n)}$ and $z_{gq}^{(n)}$, although $z_{gq}$ can be reintroduced if desired~\cite{Moch:2014sna}.  The helicity-conservation requirements lead to the transformations
\begin{eqnarray} 
\Delta I_{qq}^{(1)} &=& \Delta \tilde I_{qq}^{(1)} - z_{qq}^{(1)},
\nnu
\Delta I_{qq}^{(2,V)} &=& \Delta \tilde I_{qq}^{(2,V)}  - \Delta \tilde I_{qq}^{(1)} \otimes z_{qq}^{(1)} + z_{qq}^{(1)}\otimes z_{qq}^{(1)}
- z_{qq}^{(2,V)},
\nnu
\Delta I_{q\bar q}^{(2,V)} &=& \Delta \tilde I_{q\bar q}^{(2,V)} - z_{q\bar q}^{(2,V)},
\nnu
\Delta I_{q q}^{(2,S)} &=& \Delta \tilde I_{q q}^{(2,S)} - z_{q q}^{(2,S)},
\end{eqnarray}
We can derive 
\bea
z_{qq}^{(1)} = -8C_F(1-z)
\eea
by requiring
\bea
\Delta I_{qq}^{(1)} = I_{qq}^{(1)}.
\eea 
From helicity conservation, we can also derive $z_{qq}^{(2,V)}$ and $z_{q\bar q}^{(2,V)}$ from the $q\to q$ and $\bar q\to q$ channels, respectively.  The quantities we derive are in agreement with previous results in the literature for these quantities~\cite{Ravindran:2003gi}.  We cannot derive $z_{qq}^{(2,S)}$ from our calculation.  However, given the agreement of $z_{qq}^{(2,V)}$ and $z_{q\bar q}^{(2,V)}$ with Ref.~\cite{Ravindran:2003gi}, we take the result for $z_{qq}^{(2,S)}$ from this reference.  For completeness we list all needed terms here:
\bea
z_{qq}^{(2,V)} =& C_F^2\left[ -16(1-z) -8(2+z)\ln(z) +16(1-z) \ln(z)\ln(1-z) \right],
\nnu
&+ C_F C_A \left[\frac{4}{3}(-31+\pi^2)(1-z)-12(1+z)\ln(z)-4(1-z)\ln^2(z)\right] ,
\nnu
&-C_F \beta_0 (1-z)\left[\frac{20}{3}+4\ln(z)\right],
\nnu
z_{q\bar q}^{(2,V)} =& -C_F(C_F-C_A/2)\left[ 8(1+z)\left(4\text{Li}_2(-z) + 4 \ln(z)\ln(1+z)+2\zeta(2)-\ln^2(z) -3\ln(z)\right)-56(1-z)\right],
\\
z_{qq}^{(2,S)} =& C_F T_R \left[ 8(1-z) +4(3-z)\ln(z) +2(2+z)\ln^2(z)\right].
\eea

\bibliographystyle{h-physrev5}

\end{document}